\documentstyle[11pt]{article}
\makeatletter
\@addtoreset{equation}{section}
\newdimen\normalarrayskip
\newdimen\minarrayskip
\normalarrayskip\baselineskip
\minarrayskip\jot
\newif\ifold \oldtrue 
\def\arraymode{\ifold\relax\else\displaystyle\fi}

\def\@arrayskip{\ifold\baselineskip\z@\lineskip\z@
  \else
  \baselineskip\minarrayskip\lineskip2\minarrayskip\fi}
\def\@arrayclassz{\ifcase \@lastchclass \@acolampacol \or
\@ampacol \or \or \or \@addamp \or
 \@acolampacol \or \@firstampfalse \@acol \fi
\edef\@preamble{\@preamble
 \ifcase \@chnum
  \hfil$\relax\arraymode\@sharp$\hfil
  \or $\relax\arraymode\@sharp$\hfil
  \or \hfil$\relax\arraymode\@sharp$\fi}}
\def\@array[#1]#2{\setbox\@arstrutbox=\hbox{\vrule
  height\arraystretch \ht\strutbox
  depth\arraystretch \dp\strutbox
  width\z@}\@mkpream{#2}\edef\@preamble{\halign \noexpand\@halignto
\bgroup \tabskip\z@ \@arstrut \@preamble \tabskip\z@ \cr}%
\let\@startpbox\@@startpbox \let\@endpbox\@@endpbox
 \if #1t\vtop \else \if#1b\vbox \else \vcenter \fi\fi
 \bgroup \let\par\relax
 \let\@sharp##\let\protect\relax
 \@arrayskip\@preamble}
\@addtoreset{equation}{section}
\makeatother

\def\req#1{(\ref{#1})}

\def\commut#1#2{\left[{#1},\,{#2}\right]}

\def\frac#1#2{\mathchoice{{\textstyle{{#1}\over{#2}}}}{{#1\over#2}}%
  {{#1\over#2}}{{#1\over#2}}}

\def\d{\partial}

\def\ddl#1#2{{\stackrel{\rightarrow}{\d}\over\d #2}#1}

\def\cC{{\cal C}}

\def\cF{{\cal F}}

\def\cL{{\cal L}}
\def\cM{{\cal M}}
\def\cN{{\cal N}}

\font\frbig=eufm10 scaled\magstephalf
\font\frscr=eufm10
\font\frscrscr=eufm8
\newfam\frfam
\textfont\frfam=\frbig
\scriptfont\frfam=\frscr
\scriptscriptfont\frfam=\frscrscr
\def\fr{\fam\frfam}

\def\g{{\fr g}}

\def\PRD{Phys.\ Rev.\ D}
\def\NPB{Nucl.\ Phys.\ B}
\def\PLB{Phys.\ Lett.\ B}
\def\MPLA{Mod.\ Phys.\ Lett.\ A}
\def\CMP{Commun.\ Math.\ Phys.}

\def\JMP{J.\ Math.\ Phys.}

\newtheorem{fact}{Proposition}[section]

\newtheorem{thm}[fact]{Theorem}

\newcommand{\func}[1]{{{\cC}_{#1}}}

\newcommand{\e}[1]{\epsilon(#1)}

\begin{document}
\hfuzz=1pt
\addtolength{\baselineskip}{2pt}

\begin{flushright}
  {\tt hep-th/9901046}
\end{flushright}
\thispagestyle{empty}

\begin{center}
  {\Large{\sc A Lie Group Structure Underlying the Triplectic
      Geometry}}\\[16pt]
  {\large M.~A.~Grigoriev }\\[4pt]
  {\small\sl Tamm Theory Division, Lebedev Physics Institute, Russian
    Academy of Sciences}
\end{center}

\parbox{.9\textwidth}{\footnotesize We consider the pair of degenerate
  compatible antibrackets satisfying a generalization of the axioms
  imposed in the triplectic quantization of gauge theories.  We show
  that this actually encodes a Lie group structure, with the
  antibrackets being related to the left- and right- invariant vector
  fields on the group.  The standard triplectic quantization axioms
  then correspond to {\it Abelian\/} Lie groups.  }

\section{Introduction}
Triplectic geometry was introduced in \cite{[BMS]} (see also
\cite{[BM0],[BM2],[GS],[GS2]}) as the structure underlying the
geometrically covariant ({\it triplectic\/}) generalization of the
$Sp(2)$-symmetric Lagrangian quantization of general gauge
theories~\cite{[BLT],[BM0]}.

The most essential ingredient of the triplectic geometry is a pair of
compatible and appropriately degenerate antibrackets.  The triplectic
quantization prescription also makes use of additional objects: the
odd vector fields and a density appearing in the triplectic master
equation and path-integral.  However, the most interesting features
characteristic of the triplectic geometry originate from the
antibracket structure.

The triplectic antibrackets naturally encode two different geometrical
structures~\cite{[GS2]}.  The first is given by a complex structure
and two transversal polarizations; these {\it compatible\/} structures
are induced on the space of marked functions of the antibrackets.  The
second geometric object originating from the antibrackets is an
Abelian group structure, i.e., the structure of globally defined
nondegenerate commuting vector fields induced on the intersection
$\cL$ of the symplectic leaves of the antibrackets (i.e., on the
``manifold of fields'' in triplectic quantization).

Studied in \cite{[GS],[GS2]} were {\it mutually commutative\/}
compatible antibrackets.  The mutual commutativity condition states
that the algebra of marked functions of the antibracket $(~,~)^1$ is
commutative with respect to the antibracket $(~,~)^2$, and vice versa.
This condition is imposed for the consistency with the
$Sp(2)$-invariant quantization in the canonical
coordinates~\cite{[BLT],[BM0],[BMS]}.  As we are going to show,
however, this condition turns out to be quite restrictive from the
geometrical standpoint.  We will see that with this condition relaxed,
the compatible antibrackets induce a Lie group structure on the
submanifold~$\cL$, and conversely, any Lie group admits a ``triplectic
bundle'' endowed with a pair of compatible antibrackets.  Thus, the
construction of the ``triplectic bundle'' over a Lie group is in some
sense canonical, and a natural set of axioms may be thought to be the
one that does not lead to any further restrictions on this group.
However, the mutual commutativity axiom implies that the group is
necessarily Abelian.

This letter organized as follows: in section~\ref{sec:2}, we
introduce the basic definitions and assumptions of the triplectic
formalism and consider the simplest and most transparent example where
all the basis marked functions are Grassmann-odd and hence the
corresponding Lie group is not a {\it super\/}group.  In the
section~\ref{subsec:general}, we allow the basis marked functions to
have arbitrary Grassmann parity and construct the corresponding
submanifold that carries a Lie (super)group structure.  In
section~\ref{sec:inverse}, we explain the inverse construction: given
an arbitrary Lie group, we construct the triplectic antibrackets on a
certain bundle over it.  Finally, we make some remarks regarding the
applications to the triplectic quantization and discuss possible
generalizations of our construction.

\section{From triplectic geometry to Lie group structure} \label{sec:2}
We begin with a brief reminder of the basic structures and their
properties in the geometrically covariant approach to the
$Sp(2)$-invariant quantization.  In contrast to the geometry
underlying the covariant formulation of the standard BV formalism (see
\cite{[BT],[ASS0],[HZ],[KN],[Bering]}), the triplectic quantization
prescription requires the antibrackets to be appropriately degenerate
and to satisfy some additional constraints which we consider
momentarily.  We start, however, with the mutual commutativity
condition removed; the marked functions of each antibracket are thus
allowed to have nonzero commutators with respect to the other
antibracket.

Let $\cM$ be the ``triplectic manifold,'' i.e., a $3N$-dimensional
supermanifold endowed with a pair of compatible antibrackets.  We
assume $\cM$ to be connected (i.e., the body of $\cM$ is connected in
the standard sense).  Let $\func\cM$ be the superalgebra of smooth
functions on~$\cM$.  An antibracket on $\cM$ is a bilinear
skew-symmetric map $\func\cM \otimes \, \func\cM \to \func\cM$ satisfying
the Leibnitz rule and the Jacobi identity. A pair of antibrackets
$(~,~)^a$, where $a=1,2$, is called {\it compatible\/} if every linear
combination $(~,~)=\alpha (~,~)^1+\beta (~,~)^2$ with constant
$\alpha$ and $\beta$ is also an antibracket, i.e. satisfies the Jacobi
identity.  The compatibility condition is equivalent to
\begin{equation}
  (-1)^{(\e{F}+1)(\e{H}+1)}((F,G)^{\{a},H)^{b\}} +
  {\rm cycle}(F,G,H) =0\,,\quad
  F,G,H \in \func\cM\,,
\end{equation}
with the curly brackets denoting the symmetrization of indices $a,b$.
This condition is often referred to as the symmetrized Jacobi
identity~\cite{[BLT]}.

We take the antibrackets on $\cM$ to be everywhere of rank $2N$ and to
be {\it jointly nondegenerate\/}~\cite{[GS],[GS2]}; the latter means
that the corresponding odd Poisson bivectors do not have common zero
modes, i.e. if a 1-form $\phi$ is such that $E^1\phi=E^2\phi=0$, then
$\phi=0$, where $E^1$ and $E^2$ denote the bivectors corresponding to
the antibrackets.

We denote by $i_a\,:\,\cM_a \to \cM$ the foliations of $\cM$ into
symplectic leaves of the respective antibracket.  We also assume for
simplicity that the foliations $\cM_a \to \cM$ are fibrations, and let
$\cN_1=\cM/\cM_2$ and $\cN_2=\cM/\cM_1$ be the corresponding base
manifolds.  Then, we also have the projections $\pi_a\,:\, \cM \to
\cN_a$.

A powerful tool in studying degenerate brackets is provided by their
marked functions (the Casimir functions).  A function $\phi \in
\func\cM$ is a marked function of the antibracket $(~,~)$ if
$(\phi,F)=0$ for any $F \in \func\cM$.  An important property of the
marked functions of compatible antibrackets is as follows.
\begin{fact} \label{fact:marked}
  Let $\phi,\psi \in \func{\cM}$ be marked functions of the first
  antibracket $(~,~)^1$ {\rm(}respectively, of $(~,~)^2$ {\rm)}.  Then
  so is~$(\phi,\psi)^2$ {\rm(}respectively, $(\phi,\psi)^1${\rm)}.
\end{fact}
The proof is a direct consequence of the compatibility of the
antibrackets. It follows from the proposition that the algebra of
marked functions of the first (the second) antibracket is closed with
respect to the second (respectively, the first) bracket.

Let $\xi_{1i}$ (respectively, $\xi_{2 \alpha}$) be (locally) a minimal
set of basis marked functions of the second (respectively, the first)
antibracket.  For example, $\xi_{1i}$ can be the transversal
coordinates to the symplectic leaves of the second antibracket.  The
conditions imposed on the compatible antibrackets naturally translate
into the language of marked functions.  In particular, the rank
assumption imposed on the antibrackets implies that there are only $N$
independent marked functions of each antibracket; the nondegeneracy
condition implies that all the functions $\xi_{1i}$ and $\xi_{2
  \alpha}$ are independent.  Recalling the assumption that the
foliations $i_a\,:\,\cM_a \to \cM$ are fibrations, we see that
$\xi_{1i}$ ($\xi_{2 \alpha}$) actually constitute a local coordinate
system on $\cN_1$ (respectively, $\cN_2$).  Therefore, it is natural
to regard each $\cN_a$ as the manifold of marked functions of the
respective antibracket.

We now are in a position to study the noncommutative antibrackets.  To
avoid some technical complications, we begin with the following
instructive example.

\subsection{An instructive example}\label{subsec:example}
For simplicity, we now assume all the basis marked functions
$\xi_{1i}$ and $\xi_{2 \alpha}$ to be Grassmann-odd.  As we have
assumed the foliations of the symplectic leaves of the antibrackets to
be fibrations, there exist base manifolds $\cN_1$ and $\cN_2$, which
we further assume to be connected.  Since all the basis marked
functions are Grassmann-odd, $\cN_1$ and $\cN_2$ are odd superspaces.
The algebra $\func{\cN_1}$ ($\func{\cN_2}$) is actually a Grassmann
algebra generated by $\xi_{1i}$ (respectively, by $\xi_{2 \alpha}$).

It follows from Proposition~\ref{fact:marked} that the first
antibracket determines a skew-symmetric mapping $(~,~)^1\, :
\,\func{\cN_1} \otimes \, \func{\cN_1} \to \func{\cN_1}$ satisfying the
Jacobi identity.  In fact, the antibracket makes $\func{\cN_1}$
considered as a linear space into a Lie superalgebra, with the
antibracket being the supercommutator; here, each even element from
the Grassmann algebra~$\func{\cN_1}$ is to be considered as an odd
element of the Lie superalgebra, and vice versa.  However, we do not
discuss this in more detail and concentrate instead on some
quotient of this Lie algebra.\footnote{While this letter was in
  preparation, we received the paper~\cite{soroka}, where the
  antibracket on the Grassmann algebra corresponding to a given Lie
  algebra is considered in a slightly different context.} We
explicitly write the antibracket in $\func{\cN_1}$ as
\begin{equation}
  (\xi_{1i},\xi_{1j})^1=C^k_{ij}\xi_{1k}+\,\ldots\,,
\end{equation}
where the dots mean higher-order terms in $\xi$.  It follows from the
skew-symmetry of the antibracket that $C^k_{ij}=-C^k_{ji}$.  Moreover,
in the first order in $\xi$ the Jacobi identity for the antibracket
implies that
\begin{equation}
  C^m_{ij}C^n_{mk}+C^m_{jk}C^n_{mi}+C^m_{ki}C^n_{mj}=0\,,
\end{equation}
which is the Jacobi identity for the Lie algebra whose structure
constants are~$C^k_{ij}$.

We now explicitly construct this Lie algebra.  To this end, we
consider the vector fields on $\cM$
\begin{equation}
  L_i=(\xi_{1i},\cdot\,)^1\,,
\end{equation}
which form an $N$-dimensional Lie algebra modulo the vector fields
vanishing at $\xi_{1i}=0$,
\begin{equation}
  \commut{L_{i}}{L_j}=C^k_{ij} L_k + \ldots \,.
\end{equation}
Proceeding similarly with the algebra $\func{\cN_2}$ of the marked
functions of the first antibracket, we arrive at the vector fields
\begin{equation}
  R_\alpha=(\xi_{2\alpha},\cdot\,)^2\,,
\end{equation}
satisfying
\begin{equation}
  (\xi_{2 \alpha},\xi_{2 \beta})^2=
  C^\gamma_{\alpha \beta}\xi_{2 \gamma}+\ldots\,.
\end{equation}
In fact, $L_i$ as well as $R_\alpha$ can be restricted to the
submanifold $\cL$ determined by the equations $\xi_{1i}=0\,, \,
\xi_{2\alpha}=0$.  This allows us to consider these vector fields as
defined on $\cL$.  As a consequence of the rank conditions, $L_i$ form
a basis of $T\cL$, and so do~$R_\alpha$.  They also satisfy the
relations
\begin{equation}
  \commut{L_i}{L_j}=C^k_{ij}L_k \,, \quad
  \commut{R_\alpha}{R_\beta}=C^\gamma_{\alpha \beta} R_\gamma \,.
  \label{llrr}
\end{equation}
It follows from the compatibility of the antibrackets that
\begin{equation}
  \commut{L_i}{R_\alpha}=0\,.
  \label{LRcomm}
\end{equation}
Relations \req{llrr} and \req{LRcomm} are precisely those of the left-
and right-invariant vector fields on the Lie group corresponding to
the Lie algebra determined by the structure constants $C^k_{ij}$.  We
thus conclude that $\cL$ is diffeomorphic to $G$~\cite{[St]}.

\subsection{The general construction} \label{subsec:general}
We now turn to the general situation, where we allow the basis marked
functions to have arbitrary Grassmann parities.

As mentioned above, we identify the marked functions of the first (the
second) antibracket with functions on $\cN_2$ (respectively, on
$\cN_1$). Proposition~\ref{fact:marked} tells us that the antibracket
$(~,~)^1$ (respectively, $(~,~)^2$) induces an antibracket on $\cN_1$
(on $\cN_2$), and therefore $\cN_a\,,\,a=1,2$, become odd Poisson
manifolds.  Let $p_1 \in \cN_1$ be a point where the antibracket
$(,)^1$ vanishes,\footnote{we assume that the set of zeroes (i.e., the
  points where the corresponding odd Poisson bivector vanishes) of the
  first (the second) antibracket on $\cN_1$ (respectively, on~$\cN_2$)
  is non-empty.} i.e., $(\phi,\psi)^1|_{p_1}=0$ for all $\phi,\psi \in
\func{\cN_1}$.  This point is evidently a symplectic leaf of the
antibracket $(,)^1$ considered as an antibracket on $\cN_1$. Let
$\xi_{1i}$ be some coordinates in a neighborhood $U_{p_1} \subset
\cN_1$ of $p_1$ such that $\xi_{1i}=0$ at $p_1$. Then we have
\begin{equation}
  (\xi_{1i},\xi_{1j})^1=C^k_{ij}(\xi_1)\,\xi_{1k}\,.
  \label{struct1}
\end{equation}
with some functions $C^k_{ij}(\xi_1)$.  We will view the tensor
$C^k_{ij}=C^k_{ij}(0)$ as the structure constants of some Lie (super)
algebra $\g_1$.

Let $p_2 \in \cN_2$ be a vanishing point of the second antibracket on
$\cN_2$.  The coordinate system $\xi_{2 \alpha}$ on
$U_{p_2}\subset\cN_2$ is chosen such that $\xi_{2 \alpha}|_{p_2}=0$.
Similarly to \req{struct1}, we have the equation
\begin{equation}
  (\xi_{2 \alpha},\xi_{2 \beta})^2=
  C^\gamma_{\alpha \beta}(\xi_2)\xi_{2 \gamma}\,.
  \label{struct2}
\end{equation}
The structure constants $C^\gamma_{\alpha
  \beta}=C^\gamma_{\alpha\beta}(0)$ also give rise to a Lie algebra,
which we denote by $\g_2$.  Thus, we have associated a pair of Lie
algebras $\g_1$ and $\g_2$ to a pair of compatible antibrackets.  Note
that the Lie algebras $\g_1$ and $\g_2$ are of the same dimensions.

The vanishing points $p_a \in \cN_a$ correspond to the submanifold
$\cL \subset \cM$ determined by the equations $\xi_{1i}=\xi_{2
  \alpha}=0$, where $\xi_{1i}$ and $\xi_{2 \alpha}$ are considered as
functions on $\cM$.  In different words, $\cL= \pi_1^{-1}p_1 \cap
\pi_2^{-1}p_2$.  The functions $\xi_a$ are well-defined in some
neighborhood $U_{\cL}$ of~$\cL$ in $\cM$.  Indeed, $\xi_a$ are
functions on $\pi_a^{-1}U_{p_a}$; we then choose
$U_{\cL}=\pi_1^{-1}U_{p_1} \cap \pi_2^{-1}U_{p_2}$, which is evidently
a neighborhood containing $\cL$.  Thus the vector fields
\begin{equation}
  L_i=(\xi_{1i},\cdot\,)^1 \,,\qquad
  R_\alpha=(\xi_{2\alpha},\cdot\,)^2\,.
\end{equation}
are defined on the entire neighborhood $U_{\cL}$.  We now observe that
the vector fields restrict to $\cL$.  Indeed, it follows from
Eqs.~\req{struct1}--\req{struct2} that
\begin{equation}
  \begin{array}{c}
    (L_i \xi_{1j})|_{\cL}=(R_\alpha \xi_{2 \beta})|_{\cL}=0\,,\\
    L_i\xi_{2 \alpha}=R_{\alpha} \xi_{1j}=0 \,,
  \end{array}
\end{equation}
which are precisely the conditions for $L_i$ and $R_\alpha$ to
restrict to $\cL$.

Next, we consider $\cL$ as a submanifold in
$\cM_1=\pi^{-1}_1p_1\subset \cM$.  We have
\begin{fact}
  $\cL$ is Lagrangian submanifold of $\cM_1$.  In particular, $\cL$ is
  $N$-dimensional.
\end{fact}
We consider the functions $\xi_{1i}$ and the vector fields $L_i$ as
being defined on $\cM_1$.  The proof of the proposition follows
immediately from relations \req{struct1} and the fact that $\xi_{1i}$
are independent functions on $\cM_1$.  Since $\cM_1$ is odd symplectic
(i.e., the antibracket $(~,~)^1$ is nondegenerate on $\cM_1$) we see
that $L_i$ are linearly independent on $\cL$, and therefore form a
basis of~$T\cL$.  We can treat the vector fields $R_\alpha$ similarly,
which gives us another set of globally defined vector fields on $\cL$.
To summarize, we have
\begin{thm}\label{main}
  $\cL$ is endowed with globally defined vector fields $L_i$ and
  $R_\alpha$ that satisfy
  \begin{equation}
    \begin{array}{c}
      \commut{L_{i}}{L_j}=C^k_{ij}(0) L_k\,,\qquad
      \commut{R_\alpha}{R_\beta}=C^\gamma_{\alpha \beta}(0) R_\gamma \,,\\
      \commut{L_i}{R_\alpha}=0\,,
    \end{array}
  \end{equation}
  Therefore, $\cL$ is diffeomorphic to some Lie (super) group $G$.
\end{thm}
As a corollary of the theorem, we observe that the Lie algebras $\g_1$
and $\g_2$ are isomorphic.

Finally, we note that in the case considered in the previous
subsection, there is only one zero $p_1$ ($p_2$) of the first (the
second) antibracket on $\cN_1$ (respectively, on~$\cN_2$).  Thus,
there is only one submanifold $\cL=\pi_1^{-1} \cap \pi_2^{-1} \subset
\cM$ and the Lie group associated to the antibrackets is uniquely
determined.

In the general case, however, we have families $Z_a \subset \cN_a$ of
the vanishing points and hence a family of super Lie groups.  This
raises the question of whether different points from $Z_a$ actually
correspond to different Lie groups.  We do not address this question
here, however, and turn instead to the converse problem of
constructing triplectic antibrackets corresponding to a given Lie
group.

\section{The inverse construction}\label{sec:inverse}
The above considerations provide us with the construction of
triplectic antibrackets on the duplicated odd cotangent bundle over
any Lie group.

Let $G$ be a Lie group and $\g^*$ the corresponding Lie coalgebra.  It
is well known that the cotangent bundle $T^* G$ is trivial.  There
exist two natural ways to identify $G \times \g^*$ with $T^* G$; the
first one is to view $\g^*$ as the Lie coalgebra of left-invariant
1-forms on $G$ and the second is to view $\g^*$ as the Lie coalgebra
of right-invariant 1-forms.

Since $T^*G$ has a canonical symplectic structure, we can also equip
$G\times \g^*$ with a symplectic structure and hence with the Poisson
bracket.  We thus have canonical Poisson brackets on $G\times \g^*_l$
and on $G \times \g^*_r$, with the subscripts $r$ or $l$ indicating
the way in which we identify $G\times \g^*$ with $T^* G$.

Recall that by changing the parity of the fibres, we can associate a
supermanifold to every vector bundle.  In the present case we
associate to $T^* G$ the supermanifold $\Pi T^* G$ (with $\Pi$
denoting the parity reversing functor), which has the canonical
antibracket (the one corresponding to the canonical odd-symplectic
structure).  Thus $G \times \Pi \g^*_l$ (respectively, $G \times \Pi
\g^*_r$) is also endowed with a natural antibracket.  Let $x^A$ be a
local coordinate system on $G$ and $L_i=L_i^A\ddl{}{x^A}$ a basis of
$\g_l$ considered as the left-invariant vector fields.  Let also
$\xi_{1i}$ be the coordinates on $\Pi \g^*_l$ corresponding to the
basis dual to $L_i$.  Then the canonical antibrackets on $G \times \Pi
\g^*_l$ are
\begin{equation}
  (\xi_{1i},x^A)^1=L^A_i\,, \qquad (\xi_{1i},\xi_{1j})^1=C^k_{ij} \xi_{1k}\,,
\end{equation}
with all the other antibrackets vanishing.\footnote{A similar
  construction in the case of ordinary Poisson bracket on $G\times
  \g^*$ is well known. Its generalization to the case of antibracket
  has been considered in \cite{[AD]}} Here we have introduced the
structure constants via $\commut{L_i}{L_j}=C^k_{ij}L_k$.  In this way,
we can also obtain an explicit form of the canonical antibracket on~$G
\times \Pi \g^*_r$.

We now take the direct sum of the bundles, $\cM=(G \times \Pi \g^*_l)
\oplus (G \times \Pi \g^*_r)$.  It can be equipped with the canonical
triplectic antibrackets.
\begin{fact}
  The canonical antibrackets on $G \times \Pi \g^*_l$ and $G \times
  \Pi \g^*_r$ naturally induce a pair of compatible antibrackets on
  $\cM=(G \times \Pi \g^*_l) \oplus (G \times \Pi \g^*_r)$.  The
  antibrackets are compatible, jointly nondegenerate and satisfy the
  rank condition formulated in section~\ref{subsec:general}.
  Thus, every Lie (super)group admits a canonical triplectic bundle.
\end{fact}
Indeed, we can identify $\cM$ with $G \times \Pi \g^*_l \times \Pi
\g^*_r$; we then consider the canonical antibrackets on $G \times \Pi
\g^*_l$, which we denote by $(~,~)^1$, and the trivial (vanishing)
antibracket on $\Pi \g^*_r$.  We thus obtain the product antibracket
$(~,~)^1$ on the product $G \times \Pi \g^*_l\times \Pi \g^*_r$.  The
second antibracket $(~,~)^2$ on $G \times \Pi \g^*_l \times \Pi
\g^*_r$ is constructed similarly.

\medskip

It should be noted that in contrast to the group case, there
does not exist, in general, a natural ``triplectic'' bundle over an
arbitrary manifold $\cL$.  In fact, the construction of the
``triplectic bundle'' essentially requires the base~$\cL$ to be
parallelizable~\cite{[GS2]}.  This fact may be viewed as a serious
limitation of the triplectic quantization in its present form.  Indeed,
the triplectic antibrackets cannot be constructed (even formally)
in a covariant way in the model whose target space is a non-flat
manifold. Over a curved manifold, the triplectic structure would be
defined only locally and would depend on the choice of coordinates.
This is in contrast with the standard BV quantization~\cite{[BV]},
where the canonical antibracket exists on the odd cotangent bundle
over any field space.

A remarkable feature of our construction is that it can be repeated
for the standard (non-super) geometry.  In particular, there a exist a
pair of compatible Poisson brackets on $\cM^\prime=(G\times \g^*_l)
\oplus (G \times \g^*_r)$ for every Lie group.  Moreover, there exists
a direct analogue of Theorem~\ref{main} for a pair of appropriately
degenerate compatible Poisson brackets.

\section{Conclusions}
We have shown that the triplectic structure (consisting of a pair of
appropriately degenerate and compatible antibrackets) induces the
structure of a Lie group on the intersection of certain symplectic
leaves of the antibrackets.  The interest in the pairs of degenerate
antibrackets originates from the triplectic quantization of general
gauge theories.  In the triplectic scheme, however, one needs mutually
commutative antibrackets, and therefore the group has to be Abelian,
which appears to be a strong geometrical constraint on the
applicability of the triplectic quantization.

Our approach to the Lie group structures is somewhat reminiscent of
the well-known fact that one can naturally associate a Lie algebra to
every symplectic leaf of a Poisson bracket~\cite{[KM]}.  In the
triplectic geometry, it is not only a Lie algebra but also the
corresponding Lie group that is associated with the zero-rank
symplectic leaf of the antibracket induced on the algebra of marked
functions.  An important point of our construction is that it has a
direct analogue for the ordinary (non-super) differential geometry
based on the standard Poisson bracket.

\smallskip

We have considered the geometric structures induced on the
intersection of symplectic leaves of compatible antibrackets.  This is
only a half of the full triplectic geometry.  The other part is
concentrated on the manifold of marked functions of the antibrackets;
the corresponding geometry was studied in the case of mutually
commutative antibrackets in~\cite{[GS2]}.  This is also interesting to
generalize to the case of non-commutative antibrackets.

Another interesting aspect of the triplectic geometry is related to an
additional structure, the odd vector fields $V^a$ originating from the
one-form $\cF$ on $\cM$ \cite{[BM2]} that is compatible with the
antibrackets.  In particular, $\cF$ gives rise to a K\"ahler structure
on the manifold of marked functions provided the appropriate
nondegeneracy conditions are satisfied by~$\cF$.  This also endows
$\cL$ with an even Poisson bracket. In the case considered in this
paper, it can be checked that $\cF$ determines a left-right-invariant
Poisson bracket on $\cL$.  It seems very probable that the present
construction can be generalized to include the Poisson--Lie group
structures.

\paragraph{Acknowledgements} I am grateful to A.~M.~Semikhatov for his
attention to this work and fruitful discussions and suggestions.  I
also wish to thank I.~A.~Batalin, M.~A.~Soloviev, and I.~V.~Tyutin and
especially, O.~M.~Khudaverdian and I.~Yu.~Tipunin for illuminating
discussions.  This work was supported in part by the RFBR Grant
98-01-01155 and by the INTAS-RFBR Grant~95-0829.

\end{document}